\documentstyle[aps,epsf,pre,multicol]{revtex}
\tightenlines
\begin{document}
\title{ Persistence in higher dimensions : a finite size scaling 
study} 

\author{ G. Manoj$^{1}$ and P. Ray $^{1,*}$}

\address{ $^1$ The Institute of Mathematical Sciences, C. I. T. Campus, 
Taramani, Chennai 600 113, India \\  
$^*$ The Abdus Salam International Centre for Theoretical Physics, Trieste, Italy}

\maketitle

\begin{abstract}
We show that the persistence probability $P(t,L)$, in a coarsening system of linear size $L$ 
at a time $t$, has the finite size scaling form $P(t,L)\sim   
L^{-z\theta}f(\frac{t}{L^{z}})$ where $\theta$ is the persistence
exponent and $z$ is the coarsening exponent. 
The scaling function $f(x)\sim x^{-\theta}$ for $x \ll 1$ and is constant for large $x$.
The scaling form implies a fractal distribution of persistent sites with
power-law spatial correlations.
We study the scaling numerically for Glauber-Ising model at dimension $d = 1$ to 
4 and extend the study to the diffusion problem. Our finite size scaling 
ansatz is satisfied in all these cases providing a good estimate of the 
exponent $\theta$.   
\end{abstract}
\pacs{05.40.+j, 05.50.+q, 05.70.Ln}

\begin{multicols}{2}
Persistence decay has been the subject of considerable research activity in recent years.
The basic quantity under investigation is the persistence probability $P(t)$, which is the
probability that a given stochastic variable with zero mean retains its sign throughout
the time interval [0 : $t$]. For a large number of systems, it was found that
at asymptotic times $t$, $P(t)\sim t^{-\theta}$, where $\theta$ is in general, a
dimension dependent, non-trivial
exponent, believed to be unrelated to the other known exponents\cite{REVIEW}. 
The non-triviality of $\theta$ is particularly true for spatially extended systems
where the time evolution of the stochastic field at one lattice site is coupled to that
of its neighbours, making the effective single-site evolution non-Markovian. 

In recent times, the spatial aspects of the persistence problem has also come under
study. In particular, the spatio-temporal evolution of the set of persistent sites has
been studied by several authors. These include the diffusion problem in $d=1$\cite{ZAN},
Ising models in spatial dimension $d=1$\cite{MR0,MR1,MR2} and $d=2$\cite{JAIN} and 
the generalised $q$-state Potts model in $d=1$\cite{BRST}.
It was found that the interplay between persistence decay and the underlying coarsening
process leads to dynamical scaling and 
fractal formation in the spatial distribution of the persistent sites.
Such fractal structure has also been reported in an experimental study 
of breath figures\cite{BREATH}.

In the present paper we propose a scaling form for the persistence probability 
$P(t,L)$ as a function of the lattice size $L$ and time $t$.  
We use physical arguments to motivate the scaling form in the context of the Ising 
model and show that the scaling reflects the fractal nature and power-law 
correlations in the spatial distribution of persistent sites.  
We provide numerical evidence for its validity through simulations in
spatial dimension $d=1$ to 4. The analysis is further
extended to the diffusion problem where approximate analytic theories have
been used to predict $\theta$ in all dimensions.
We argue that fractal formation in diffusion should take place 
in all dimensions and provide supportive results from simulations. 

Let us consider the Ising model
in a $d$-dimensional geometry of linear size $L$. We start from an initial random
configuration and quench the system, say, to the temperature $T=0$. 
As a result, the spins 
evolve in time following the Glauber dynamics, lowering the total 
energy of the configuration in
the process. In course of time, domains of positive and negative spins form, with
characteristic length scale $\xi(t)$ growing as a power-law in time ie., $\xi(t)\sim t^{1/z}$,
where $z$ is the dynamical exponent for the coarsening process\cite{PHOD}. 
The fraction of persistent
spins decays as power of time : 
$P(t,L)\sim t^{-\theta}$ as long as $t\ll t^{*} \sim L^{z}$.   
For $t \gg t^{*}$, the domain cannot grow any further because of the finite system size 
and persistence probability stops decaying, attaining a limiting value   
$P(\infty,L)\sim L^{-z\theta}$. This happens as long as 

\begin{equation}
\frac{z\theta}{d}< 1  
\label{eq:FRAC}
\end{equation}

For $z\theta > d$, persistence probability will decay to zero for
any lattice size $L$. Also we assume that there is no `blocking', whereby a finite
fraction of spins never flip, leading to a limiting value $P_{\infty}$ independent
of finite size effects. Such a situation is believed to occur in Ising model for dimensions
$d > 4$\cite{STAUFFER} and in disordered systems \cite{CMNEWMAN}.

The above behaviour of the persistent
fraction $P(t,L)$ for finite lattice sizes can be summarised in the following dynamical
scaling form.

\begin{equation}
P(t,L)=L^{-z\theta}f(t/L^{z})
\label{eq:SCALFORM}
\end{equation}

where the scaling function $f(x)\sim x^{-\theta}$ for $x \ll 1$ and $f(x)\to$ constant 
at large $x$. Similar finite size scaling ideas have been used in a previous
work in the context of global persistence exponent for nonequilibrium critical
dynamics\cite{SATYA1}.

The finite-size scaling form given by Eq.\ref{eq:SCALFORM}
implies the presence of scale-invariant spatial correlations in the system, characteristic
of fractals. To show this, we consider the two-point
correlation function $C(r,t)$, which we define as the probability of finding a 
persistent spin at a distance $r$ from another persistent spin. 
For a $d$-dimensional system, $C(r,t)$ satisfies the normalisation condition
$\int_{0}^{L}C(r,t)d^{d}r=L^{d}P(t,L)$. After substituting 
Eq.\ref{eq:SCALFORM}, this becomes

\begin{equation}
\int_{0}^{L}C(r,t)r^{d-1}dr\sim L^{d-z\theta}f(t/L^{z})
\end{equation}

Let us rewrite this equation in terms of 
a new function $F(a,b)=a^{z\theta}C(a,b)$ and 
dimensionless variables $x=r/L$ and $\tau=t/L^{z}$.

\begin{equation}
\int_{0}^{1}F(Lx,L^{z}\tau)x^{d-1-z\theta}dx\sim f(\tau)
\end{equation}

Since the RHS of the equation has no explicit $L$-dependence, LHS should also
be likewise. This is possible only if $F(a,b)=g(ba^{-z})$, where $g(\eta)$ is
given by the integral relation

\begin{equation}
\tau^{\frac{d}{z}-\theta}\int_{\tau}^{\infty}\eta^{\theta-(1+\frac{d}{z})}g(\eta)d\eta
\sim zf(\tau)
\label{eq:INTEGRAL}
\end{equation}

Using the above equation, the limiting behaviour of the function $g(\eta)$ for small
and large values of the argument could be deduced from the known behaviour
of the function $f(\tau)$. Consider $\tau\gg 1$, where $f(\tau)$ is constant.
From Eq. \ref{eq:INTEGRAL}, this implies that $g(\eta)$ is constant for large $\eta$.
In the other extreme of $\tau\ll 1$, $f(\tau)\sim \tau^{-\theta}$. We split    
the integral in Eq. \ref{eq:INTEGRAL} as $\int_{\tau}^{\infty}=
\int_{\tau}^{\alpha} + \int_{\alpha}^{\infty}$  and note that $g(\eta)$ is   
constant in the second integral for sufficiently large $\alpha$.   
The second integral
vanishes as $\tau^{\frac{d}{z}-\theta}$ as $\tau\to 0$, whereas the RHS diverges as
$\tau^{-\theta}$. This can be consistent only if the first integral diverges as   
$\tau^{-\theta}$, which would imply that $g(\eta)\sim \eta^{-\theta}$ as $\eta\to 0$.
This leads to the following dynamical scaling form for $C(r,t)$.

\begin{equation} 
C(r,t)=r^{-z\theta}g(\frac{t}{r^{z}})
\label{eq:CORR}
\end{equation}

For small separations $r\ll t^{1/z}$, this scaling form implies 
scale-free correlations, ie., $C(r,t)\sim r^{-z\theta}$, characteristic of a fractal
with fractal dimension $d_{f}=d-z\theta$. On the other hand, over larger length scales,
$C(r,t)\sim t^{-\theta}$, which is indicative of the absence of any spatial
correlations. This scaling description was introduced by us\cite{MR1,MR2} 
in the context of $A+A\to\emptyset$ model, and later verified 
numerically in 2-dimensional Ising model\cite{JAIN} also.

To check the finite-size scaling form given by Eq. \ref{eq:SCALFORM}, 
we simulate Ising spin
systems of various sizes in spatial dimension $d=1$ to 4. Starting from a
random initial configuration, the spins are quenched to zero temperature and are 
updated sequentially using the Glauber updating rule by which  
a spin is always flipped if the resulting energy change $\Delta E < 0$, never flipped if
$\Delta E >0$, and flipped with probability $\frac{1}{2}$ if $\Delta E=0$. 
One MC time step was counted
after every spin in the lattice was updated once.  
The persistence probability at any time $t$ was determined as the fraction
of spins that did not flip even once till time $t$ since the time evolution started. 
The data is averaged typically over 1000 starting random configurations for small 
$L$ and low $d$ and over $50$ starting configurations for large $L$ and 
high $d$. 

\vspace{-0.5cm}
\begin{figure}[a]
\narrowtext
\epsfxsize=1.7in
\hspace{-1.5cm}
\epsfbox{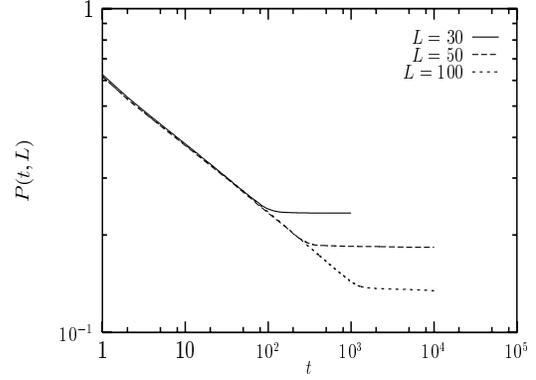}
\vspace{-0.5cm}
\caption{The persistence probability $P(t,L)$ is plotted against time $t$  
(measured in MC steps) for three different lattice sizes $L$ in $d=2$ Glauber Ising model.}
\end{figure}

\vspace{-0.5cm}
\begin{figure}[b]
\narrowtext
\epsfxsize=1.7in
\hspace{-1.5cm}
\epsfbox{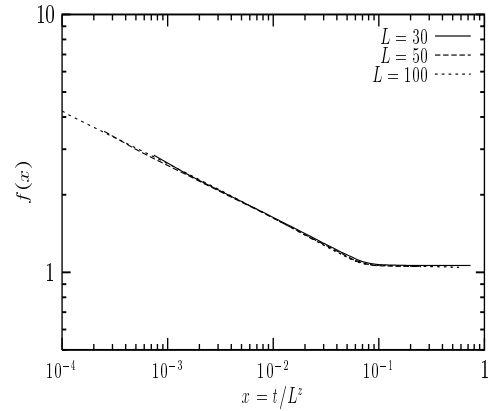}
\vspace{-0.5cm}
\caption{
Same as Fig. 1, except that the scaling function $f(x)=L^{z\theta}P(t,L)$
is plotted against the dimensionless scaling variable $x=t/L^{z}$. The data for different
$L$ values were found to collapse well to a single curve for $\theta=0.21$
and $z=2.12\pm 0.05$.  
}
\end{figure}

For $T=0$ Glauber dynamics of Ising model, the persistence exponent $\theta$ is 
exactly
known to be $3/8$ in $d=1$\cite{DERRIDA}. In higher dimensions, simulations predict
$\theta\simeq 0.22(d=2)$\cite{STAUFFER,DERRIDA1,SATYA} and $\theta\simeq 0.16 (d=3)$
\cite{STAUFFER}. 
In our finite size scaling analysis of the simulation data, we adopt the following
procedure. For $d=1,2$ and 3, we fix $\theta$ at its known value and adjust
$z$ to find the value which gives the best data collapse. In all cases, we find
$z\simeq 2$, which is the accepted value of the coarsening exponent for
non-conserved scalar models\cite{PHOD}. 
(In $d=3$ Galuber dynamics, a slower $t^{1/3}$ coarsening has been
observed before\cite{SETHNA}. This is presumably due to 
lattice effects, but we have not seen any signature of this effect 
in our simulations).
In $d=4$, on the other hand,
we fix $z$ at 2, and adjust $\theta$ to collapse the data to a single curve.
The results are displayed in Figs. 1 to 4.


In $d=4$, we find that for $z=2$, $\theta=0.12\pm 0.02$ gives reasonably good data 
collapse over the time scales and system sizes studied. Fig.4 shows the 
scaled data in $d=4$. It may be mentioned that in $d=4,$ earlier simulations 
had suggested that the persistence decay might be slower than a power-law, 
and perhaps logarithmic\cite{STAUFFER}. However, the agreement of our data
with the scaling form Eq.\ref{eq:SCALFORM} suggests that persistence follows
a power-law decay in $d=4$ also. For $d > 4 $, blocking of spins has been
shown to lead to a limiting value of $P(t,L)$ as $t\to\infty$, which is
independent of $L$\cite{STAUFFER}. We could simulate only small lattice sizes for 
$d=5$ from which we cannot make any conclusive remark at this stage.

\vspace{-0.5cm}
\begin{figure}[c]
\narrowtext
\epsfxsize=1.7in
\hspace{-1.5cm}
\epsfbox{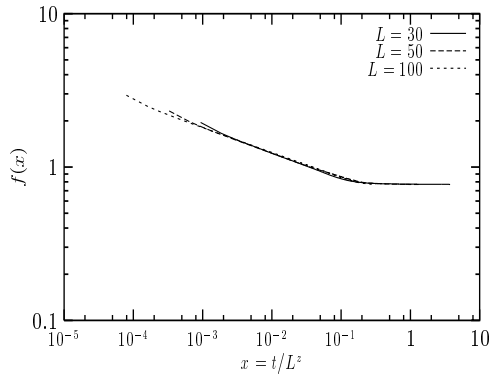}
\vspace{-0.5cm}
\caption{The scaling function $f(x)=L^{z\theta}P(t,L)$ is plotted
against the dimensionless scaled time $x=t/L^{z}$ for three $L$-values in $d=3$ Glauber 
Ising model. 
The observed data collapse has been obtained for $z=2.05$
and $\theta=0.166$.} 
\end{figure}

\vspace{-0.5cm}
\begin{figure}[d]
\narrowtext
\epsfxsize=1.7in
\hspace{-1.5cm}
\epsfbox{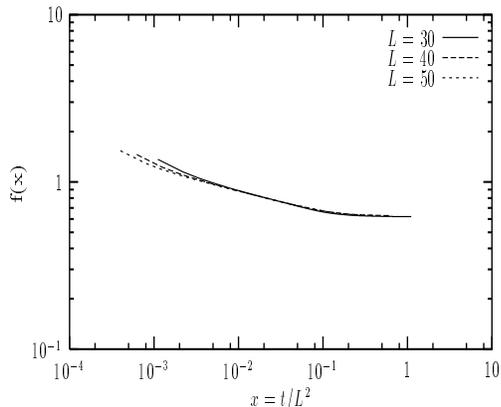}
\vspace{-0.5cm}
\caption{ The figure shows the scaled probability plotted against the
dimensionless scaled time in $d=4$ Glauber Ising model. We have fixed $z=2$, and find that 
$\theta=0.12\pm 0.02$ gives the best data collapse.} 
\end{figure}

In the diffusion problem, we have a scalar field $\phi({\bf x},t)$ evolving according to
the diffusion equation. The initial values $\phi({\bf x},0)$ are taken to be independent
random variables with zero mean.

\begin{equation} 
\frac{\partial \phi({\bf x},t)}{\partial t}={\nabla}^{2}\phi({\bf x},t)
\hspace{0.3cm};\hspace{0.3cm}\langle\phi({\bf x},0)\phi({\bf x}^{\prime},0)\rangle=
\delta^{d}({\bf x}-{\bf x}^{\prime})
\label{eq:DIFF}
\end{equation}

For this problem, it has been shown using approximate analytic 
theories\cite{DIFF1,DIFF2,DIFF3}that $P(t)\sim t^{-\theta}$ in 
all dimensions. The predicted exponent values in low dimensions were 
in good agreement with simulation results. 
The exponent was found to increase with dimension, and has been suggested 
to have the asymptotic value 
$\theta(d)\simeq \alpha\sqrt{d}$ as $d\to\infty$. The constant $\alpha$ 
has been estimated
to be $\simeq 0.14$\cite{DIFF1,DIFF2} and $\simeq 0.18$\cite{DIFF3} by different authors. 
For $d=1,2$ and 3, the exponent values are found to be $\theta \simeq 0.12, 0.18$ and $0.23$ 
respectively. 

To simulate Eq. \ref{eq:DIFF} numerically, we use the finite difference Euler discretization
scheme on cubic lattices of $L^{d}$ sites \cite{DIFF1,DIFF2}.

\begin{equation}
\phi({\bf x},t+\Delta t)=\phi({\bf x},t)+a\left[\sum_{{\bf x}^{\prime}}\phi({\bf
x}^{\prime},t)-2d\phi({\bf x},t)\right]
\label{eq:DISCRETE}
\end{equation}

where ${\bf x}^{\prime}$ runs over all the $2d$ nearest neighbour lattice sites
of ${\bf x}$ in the cubic lattice and $a=\frac{\Delta t}{(\Delta x)^{2}} < 
\frac{1}{2d}$
for stability of the discretization scheme. We have taken $a=\frac{1}{4d}$ in our
simulations as this value has been observed to provide the fastest approach to
the asymptotic regime\cite{DIFF1}. 

For the diffusion problem, simple scaling arguments suggest that the dynamical 
exponent $z=2$ in all dimensions. In all dimensions studied,
we found excellent scaling collapse with $z\simeq 2$ and the $\theta$ values quoted
above. Upon substitution of the
exponent values into Eq. \ref{eq:FRAC}, it can be easily seen that the condition for fractal
formation is satisfied for $d=1,2$ and 3. For $d=1$, this has already been confirmed 
by an earlier numerical study\cite{ZAN}. Our results for the persistence probability
and the scaling function for three different lattice sizes 
in $d=2$ is displayed in Fig. 5 and 6.

It is also possible to extrapolate these results to the $d\to\infty$ limit using the
asymptotic form suggested for $\theta$. We see that in this limit, the LHS of Eq. \ref{eq:FRAC}
vanishes as $\frac{1}{\sqrt{d}}$, leading us to conjecture that fractal formation persists
in all dimensions for the diffusion problem.

\vspace{0.5cm}
\begin{figure}[e]
\narrowtext
\epsfxsize=1.7in
\hspace{-1.5cm}
\epsfbox{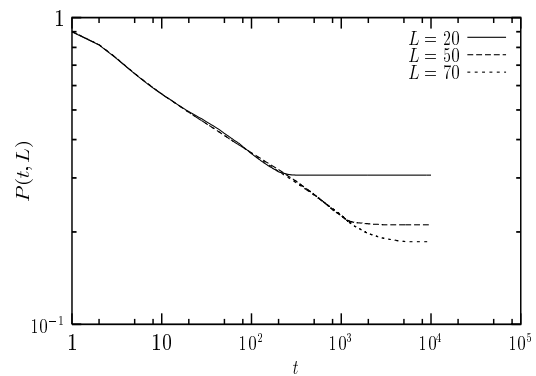}
\vspace{-0.5cm}
\caption{The persistence probability $P(t,L)$ is plotted against time $t$  
(measured as the number of MC steps) for three different lattice sizes $L$ in $d=2$ diffusion problem.}
\end{figure}

\vspace{-0.5cm}
\begin{figure}[f]
\narrowtext
\epsfxsize=1.7in
\hspace{-1.5cm}
\epsfbox{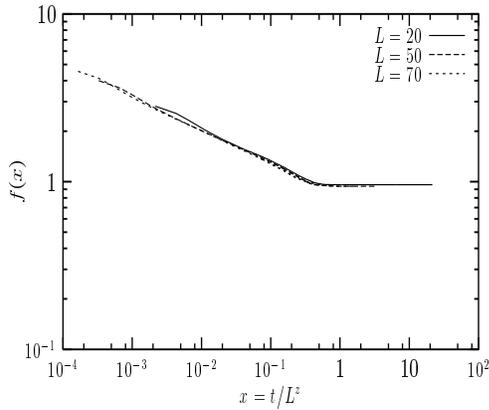}
\vspace{-0.5cm}
\caption{
Same as Fig. 5, except that the scaling function $f(x)=L^{z\theta}P(t,L)$
is plotted against the dimensionless scaling variable $x=t/L^{z}$. The data for different
$L$ values were found to collapse well to a single curve for $\theta=0.186$
and $z=2.05\pm 0.04$.  
}
\end{figure}

To conclude, we have proposed a finite size scaling ansatz for the persistence
probability in a coarsening system. The scaling form corresponds to the fractal
structure and dynamic scaling characterising the spatio-temporal evolution
of the persistent set. We check the scaling form numerically for Glauber-Ising model
and for the diffusion problem. Finite size scaling enables us to study persistence
reliably in higher dimensions. Our results agree with the known values of $\theta$ in 
the case of Ising model(from $d=1$ to 3) and in the diffusion problem (we have checked
upto $d=3$). For $d=4$ Ising model, we find the signature of algebraic decay of persistence
with $\theta\simeq 0.12$, in contrast with what had been reported earlier\cite{STAUFFER}. 

We thank G. I. Menon and D. Dhar for a critical reading of the manuscript 
and valuable suggestions.
G.M gratefully acknowledges the hospitality at The Abdus Salam ICTP, Trieste, 
Italy where this work was done.
G. M also thanks C. Sire, S. N. Majumdar and A. J. Bray for helpful discussions 
and illuminating remarks.  


\end{multicols}

\end{document}